\documentclass[preprint]{aastex}
 
\newcommand{\be}{\begin{equation}} 
\newcommand{\ee}{\end{equation}}

\newcommand{\bea}{\begin{eqnarray}} 
\newcommand{\eea}{\end{eqnarray}} 
 
\newcommand{\ie}{i.e., } 
\newcommand{\OM}{\Omega_m} 
\newcommand{\OO}{\Omega_0}
\newcommand{\OL}{\Omega_{\Lambda}}

\begin{document}
\title{Distance-Redshift in Inhomogeneous FLRW} 
\author{ R.  Kantowski }
\affil{ University of Oklahoma, Department of Physics and
Astronomy,\\ Norman, OK 73019, USA }
\email{kantowski@mail.nhn.ou.edu}

\author{ J. K. Kao}
\affil{ Tamkang University, Department of Physics,\\
Tamsui, Taipei, Taiwan 25137 R.O.C.}
\email{g3180011@tkgis.tku.edu.tw}
\author{ R. C. Thomas }
\affil{ University of Oklahoma, Department of Physics and
Astronomy,\\ Norman, OK 73019, USA }
\email{thomas@mail.nhn.ou.edu}

\begin{abstract} \vskip .2 truein We give distance--redshift relations 
in terms of elliptic integrals
for three different mass distributions of
the Friedmann-Lema\^\i tre-Robertson-Walker (FLRW) 
cosmology. These models are dynamically pressure free FLRW on large scales
but, due to mass inhomogeneities, differ in their optical properties. 
They are the filled-beam model 
(standard FLRW), the empty-beam model (no mass density exists in the 
observing beams) and the 2/3 filled-beam model. For special $\OM$--\ $\OL$ values
the elliptic integrals reduce to more familiar functions. These new expressions 
for distance-redshift significantly reduce computer evaluation times.
\end{abstract}


\section{INTRODUCTION} \label{sec-intro} 
As limits on the global cosmological parameters $\OM$ and $\Lambda$ have been refined,
\cite{SB} and \cite{PS1},
the optical inadequacy of the standard distance-redshift relation ($D$-$z$) of 
FLRW has become more apparent. The problem was first recognized long ago by 
\cite{Zel}, \cite{BB}, and \cite{KR}
but the lack of relevant data limited its significance. Even though the average mass density 
parameter $\OM$ (along with $H_0$ and $\Lambda$) determines the large scale dynamic behavior 
of the pressure free universe, knowledge of the actual mass inhomogeneity is necessary to 
accurately determine these 
parameters from most observations. Most observations determine $\OM$ and $\Lambda$ by 
(indirectly) comparing theoretical $D$-$z$ curves to observed data. However, 
 $D$-$z$ depends on more than the average mass density. It can depend significantly 
on details of how the mass is distributed, \ie on how inhomogeneous the mass is on
the scale of the widths of the observing beams.  
If some significant fraction ($\rho_I/\rho_0 \le 1$) 
of the total mass density is in the form of inhomogeneities and is excluded from 
the lines of sight to the distant objects observed, a modified, \ie a partially 
filled-beam $D$-$z$ is  required. 

The necessity of taking into account the effect of inhomogeneities on 
observations is relatively easy to understand. 
Homogeneous matter
inside an observing beam of light gravitationally focuses
the beam much differently than does an equal-mass clump of externally lensing matter. 
The simplest correction for this gravity-light effect requires the introduction 
of another parameter $\nu,\ 
0\le\nu\le 2,$ which gives the fraction $\rho_I/\rho_0 =\nu(\nu+1)/6$ of the mass density
of the universe removed from the observing beams as inhomogeneities. Using $\nu$ 
rather than  $\rho_I/\rho_0$ or some other parameter is dictated by the mathematics of 
special functions. 
A reduced mass density in an observing beam causes it to diverge relative
to a standard FLRW beam. For an observed object in such a universe 
to have the standard FLRW 
angular size it would thus have to be moved to a smaller $z$; i.e., objects will appear
less bright than in the standard FLRW universe.
A reasonable application of this model to SNe Ia observations takes  
$\rho_I$ as the galactic contribution to the total mass density $\rho_0$ and the remaining 
contribution as a smooth intergalactic medium. 
Galaxies are easily excluded from SNe Ia foregrounds by selection (intended or not)
and if galaxy mass roughly follows light, including their mass in $\rho_I$ is appropriate.
In the partially filled-beam model where the additional  parameter $\nu\ne 0$ 
has been introduced,  
only lensing by mass clumps external to the beam has been neglected. 
To compare individual observations to  $D$-$z$ 
of this model  requires only  an occasional lensing correction; however, 
comparison with the standard FLRW $D$-$z$ ($\nu=0$) model requires a defocusing correction
for the partially empty-beam of every observation, as well as the 
occasional lensing  correction. 
If only weak and transparent lensing occurs (to the $z_{max}$ being observed)
the standard  FLRW $D$-$z$ ($\nu=0$) should give the mean $D$-$z$ curve. 
\cite{WY} argues that by using flux-averaging the mean can be accurately obtained.
\cite{KRa} and \cite{KRb} claims that determining cosmological parameters 
 from data compared 
with the partially filled Hubble curves given here is likely to be easier. 
Beyond selection effects, 
unknown lensing probabilities can be highly non-Gaussian and should make the mean more 
difficult 
to observationally determine, \ie should require more data if a 
given accuracy of the cosmic parameters is to be obtained, \cite{BB,HW,HD}. 
The down side for partially filled-beam models is that you must select against
lensing and must determine the additional parameter $\nu$.

In Sec.\,\ref{sec-lumdist} we outline the procedure required to obtain $D$-$z$
for partially filled-beam FLRW observations and how the result simplifies for the 
three special cases of $\nu =$ 0, 1, and 2. In Sec.\,\ref{sec-results} we give the 
new results for these three special cases. Some concluding remarks are given in 
Sec.\,\ref{sec-conclusions} and in the Appendix we discuss our Fortran 
implementation of these results.

\section{The Luminosity Distance-redshift Relation} \label{sec-lumdist} 

For models being discussed here (and for most cosmological models), 
angular or apparent size distance 
is related to luminosity distance by $D_<(z)=D_{\ell}(z)/(1+z)^2$. 
Hence we need to give only one
or the other, and we have chosen to give luminosity distances. 
The $D_{\ell}(z)$ which accounts
for a partially depleted mass density in the observing beam but neglects 
lensing by external masses
is found by integrating the second order differential equation for the 
cross sectional area $A(z)$
 of an observing beam from source ($z=z_s$) to observer ($z=0$), see \cite{KRa}
for some history of this equation:
\bea 
&&(1+z)^3\sqrt{1+\OM z+\OL[(1+z)^{-2}-1]}\times\nonumber\\ &&\hskip 1 in {d\ \over
dz}(1+z)^3\sqrt{1+\OM z+ \OL[(1+z)^{-2}-1]}\,{d\ \over dz}\sqrt{A(z)}\nonumber\\ &&\hskip 2.0 in
+ {(3+\nu)(2-\nu)\over 4}\OM(1+z)^5\sqrt{A(z)}=0.   \label{lame} 
\eea
The required  boundary conditions are:
\bea
\sqrt{A}|_s&=&0,\nonumber\\ {d\sqrt{A\big|_s }\over dz}&=& -\sqrt{\delta\Omega} {c\over
H_s(1+z_s)}, \label{Aboundary} 
\eea 
 where $\delta\Omega$ is the solid angle of the beam at the source
and 
the FLRW value of the Hubble parameter at $z_s$ is
related to the current value $H_0$ at $z=0$ by: 
\be
H_s=H_0(1+z_s)\ \sqrt{1+\OM
z_s+\OL[(1+z_s)^{-2}-1]}.  \label{Hs} 
\ee
The luminosity distance is then simply related to the area $A\big|_0$ of the beam 
at the observer by: 
\be
D_{\ell}^2\equiv {A\big|_0 \over \delta\Omega}(1+z_s)^2.
\label{Dl} 
\ee
Equation (\ref{lame}) can be put into the form of a Lam\'e equation
and its solution has been given in terms of Heun 
functions in \cite{KRa}. Solutions can also be given in terms of Lam\'e functions 
but neither Heun nor Lam\'e functions are currently available 
in standard computer libraries. Consequently, such expressions are not particularly 
useful for comparison with data, at this time. 
For the special case where $\Lambda=0$ the Lam\'e functions reduce to 
associated Legendre functions and these expressions are useful. Other special 
cases also exist as is pointed out in \cite{KRa}.

In the next section we give useful expressions for 
$D_{\ell}$ for three special cases where  $\Lambda$ is arbitrary but 
where the filling parameter $\nu$ is restricted to values 0, 1, and 2. For these 
three cases we can write $D_{\ell}$ as an elliptic integral and hence we can
give  $D_{\ell}$ in terms of the three fundamental incomplete Legendre elliptic integrals
$F(\phi,{\rm k}), E(\phi,{\rm k}),$ and $ \Pi(\phi,\alpha^2,{\rm k})$. These functions are universally
available and these new expressions significantly speed up the evaluation of 
$D_{\ell}$\ (see the Appendix). Distance-redshift for $\OO=1$ 
can be given in terms of hypergeometric functions, see (\ref{2F1ansnu=0B1}) and (\ref{2F1ansnu=2B1}), 
or associated Legendre functions, 
see (\ref{Pansnu=0B1}) and (\ref{Pansnu=2B1}); however, we also give $D_{\ell}$ as 
more complicated expressions
involving Legendre elliptic integrals, (\ref{ansnu=0B1}) and (\ref{ansnu=2B1}), 
because these 
expressions evaluate more rapidly using currently available Fortran routines.

It is not at all clear that the solution of (\ref{lame}) can be written as elliptic integrals
for the special cases of $\nu=$ 0, 1 and 2. However, the steps required to arrive at this 
conclusion
can be found in \cite{WE} under integral functions for Lam\'e and Matthew equations
(see especially Sec.\,19.53). The authors have carried out the 
conversion directly for all three cases; however, the $\nu=0$ and $2$ conversions can be reached 
by simpler means. The integral for  $\nu=0$, the standard FLRW filled-beam case, 
is given in (\ref{nu=0}) and is well known.
The $\nu=2$ (empty-beam) integral given in (\ref{nu=2}) is easy to obtain because the 
coefficient 
of $\sqrt{A}$ vanishes in (\ref{lame}). The first integral is trivial and the second
is elliptic resulting in (\ref{nu=2}).
For $\nu=1$, the 66\% 
filled-beam model, the integral is given in (\ref{ansnu=1A}); however, no simple way of getting this from 
 (\ref{lame}) seems to exist. 

In Sec. \ref{sec-results}. we outline results for all big bang models in the first 
quadrant of the $\OM$--\ $\OL$ plane (see Fig. 1), hoping to facilitate their usage. 
Luminosity distances for the three large open domains are given in subsections A, and 
for the boundaries of these domains in subsections B.
\section{Luminosity Distances as Legendre Elliptic Integrals} \label{sec-results}

\centerline{\bf I. $\nu=0$, Completely Filled-Beam Observations (Standard FLRW)}
{\bf A. Three Open Big Bang Domains}

\cite{KSSE} and \cite{KS} gave magnitude-redshift relations for standard pressure-free FLRW 
models as 
inverse  Weierstrass functions and more recently \cite{FB} gave
comoving distances and light travel times for these models using Legendre elliptic integrals. 
In this section we give simpler and more useful results which are directly comparable with 
\cite{ED} who used Jacobi elliptic functions. 
The well known and often used integral form for luminosity distance in standard FLRW is:
\be 
D_{\ell}(\OM,\OL,\nu=0;z)= {c\over H_0}{1+z \over \sqrt{|1-\OO|}}\ S_{\kappa}
\left[ \sqrt{|1-\OO|} 
\int_0^z{dz\over \sqrt{(1+z)^2(1+\OM z)-z(z+2)\OL}}
\right]\label{nu=0}
\ee
which we integrate using \cite{BF} to obtain,
\be
D_{\ell}(\OM,\OL,\nu=0;z)={c\over H_0}{1+z \over \sqrt{|1-\OO|}}\ S_{\kappa}
\Bigl[\ 
-g
\Bigl\{
F(\phi_z,{\rm k})-F(\phi_0,{\rm k})
\Bigr\}
\Bigr],
\label{ansnu=0}
\ee
or equivalently using an addition formula for $F(\phi,{\rm k})$, \ie 
$F(\phi_z,{\rm k})-F(\phi_0,{\rm k})$ = $F(\Delta\phi_z,{\rm k})$ we get:
\be
D_{\ell}(\OM,\OL,\nu=0;z)={c\over H_0}{1+z \over \sqrt{|1-\OO|}}\ S_{\kappa}
\Bigl[\ 
-g\
F(\Delta\phi_z,{\rm k})
\Bigr].
\label{Del_ansnu=0}
\ee
The parameter  $\kappa\equiv$ ($\OO-1$)/$|\OO-1|$ is determined by the sign of the 
3-curvature and $S_{\kappa}[\ ]$ is 
one of two functions:
\[
S_{\kappa}[\ ]= \left\{
\begin{array}{l c  l}
{\rm sinh}[\ ] &:& \kappa= -1, \\
{\rm sin}[\ ] &:& \kappa = +1.
\end{array}
\right.
\]
Constants $g$ and ${\rm k}$  depend on the cosmic parameters $\OM\ \&\ \OL$, and $F(\phi,{\rm k})$ 
is the incomplete Legendre 
elliptic integral of the first kind.\footnote{
$F(\phi,{\rm k})\equiv \int_0^{\phi} 1/\sqrt{1-{\rm k}^2\sin^2\phi}\ d\phi$}
The constants $g$ and ${\rm k}$ depend on $\OM\ \&\ \OL$
only through a  combination called $b$ defined by:
\be
b\equiv -(27/2){\OM^2\OL\over (1-\OO)^3}, \hskip .5 in -\infty\le b \le \infty,
\ee
\bea
b<0 \ \Leftrightarrow \kappa = -1,\nonumber\\
b>0 \ \Leftrightarrow \kappa = +1.\nonumber
\eea
The functions $\phi_z$ and $\Delta\phi_z$ depend on the redshift $z$ and the cosmic parameters 
$\OM\ \&\ \OL$ (not just on the combination $b$). Domains for the various $b$ 
values in the $\OM$--\ $\OL$ plane are shown in Fig. 1.

{\bf 1.} For the two open domains defined by $b < 0$ and $2 < b$, quantities 
$g,{\rm k},$ $\phi_z$, and  $\Delta\phi_z$ are conveniently written in 
terms of intermediate constants $v_{\kappa},\ y_1$ and $A$
defined by:
\be  
v_{\kappa}\equiv\left[\kappa(b-1)+\sqrt{b(b-2)}\right]^{1/3}, \hskip .5in v_{\kappa}\ge 1.
\label{v}
\ee 
\be
y_1\equiv{-1+\kappa (v_{\kappa}+v_{\kappa}^{-1})\over 3},
\label{y1kappa}
\ee
\be
A=A(\OM,\OL)\equiv \sqrt{y_1(3y_1+2)}=\sqrt{{v_{\kappa}^2+v_{\kappa}^{-2}+1\over 3}}\ge 1.
\label{A}
\ee
Parameters $g$ and ${\rm k}$ are then given by:
\be
g=g(\OM,\OL)= 1/\sqrt{A(\OM,\OL)}=
\left[ {3\over v_{\kappa}^2+v_{\kappa}^{-2}+1}\right]^{1/4}\le 1,
\ee
and
\be
{\rm k}^2={\rm k}^2(\OM,\OL)= {2A+\kappa(1+3y_1)\over 4A}= 
\left[{1\over 2}+{1\over 4}g^2(v_{\kappa}+v_{\kappa}^{-1})\right]\le 1.
\ee
Functions $\phi_z$, and  $\Delta\phi_z$ are given by:
\be
\phi_z=\phi(\OM,\OL;z)
=\cos^{-1}
\left[
{(1+z)\OM/|1-\OO|+\kappa y_1-A
\over
(1+z)\OM/|1-\OO|+\kappa y_1+A}
\right],
\label{phiA1}
\ee
and
\be
\Delta\phi_z=\Delta\phi(\OM,\OL;z)
= 2\ \tan^{-1}\left[{-z\ \sqrt{A}\sqrt{|1-\OO|}\sqrt{1+z[1-(1-\OO)\OM^{-1}y_1]^{-1}}
\over
1+z [1-(1-\OO)\OM^{-1}y_1]^{-1}+\sqrt{(1+z)^2(1+\OM z)-z(z+2)\OL}}
\right].
\label{Del_phiA1}
\ee

{\bf 2.} For the domain $0 < b < 2$ ($\Rightarrow \kappa=1$) three intermediate parameter 
$y_1,\ y_2$ and
$y_3$ are 
convenient to use, although none are really necessary. In 
this domain of $b$, intermediate parameters $y_1,\ y_2$  and $y_2$ are related to the cosmic parameters 
$\OM \&\ \OL$ through $b$ by:
\bea
y_1&\equiv& {1\over 3}
\left(
-1+{\rm cos}\left[{{\rm cos}^{-1}(1-b)\over 3}\right]+
\sqrt{3}\ {\rm sin}\left[{{\rm cos}^{-1}(1-b)\over 3}\right]\right), \ \ 0\le y_1\le 1/3,
\nonumber\\
y_2&\equiv& {1\over 3}
\left(
-1-2\ {\rm cos}\left[{{\rm cos}^{-1}(1-b)\over 3}\right]\right), \ \ -1\le y_2\le -2/3,
\nonumber\\
y_3&\equiv& {1\over 3}
\left(
-1+{\rm cos}\left[{{\rm cos}^{-1}(1-b)\over 3}\right]-
\sqrt{3}\ {\rm sin}\left[{{\rm cos}^{-1}(1-b)\over 3}\right]\right), \ \ -2/3\le y_3\le 0.
\label{y123}
\eea
The following expressions are valid 
{\bf only} in the lower right part of the $\OM$--\ $\OL$ plane. 
In the upper left domain where $b$ also satisfies $0\le b \le 2$,
expressions can be given, but
there a big bang 
doesn't occur. The parameters $g$ and ${\rm k}$ and functions $\phi_z$ and $\Delta\phi_z$ 
needed to evaluate (\ref{ansnu=0}) and (\ref{Del_ansnu=0}) are:
\be
g=g(\OM,\OL)\equiv {2\over \sqrt{y_1-y_2}},
\ee
\be
{\rm k}^2={\rm k}^2(\OM,\OL)\equiv { y_1-y_3\over y_1-y_2 }\le 1,
\label{k+}
\ee
\be
\phi_z=\phi(\OM,\OL;z)={\rm sin}^{-1}
\sqrt{
y_1-y_2
\over
(1+z)\OM/|1-\OO|+y_1
}
\label{phiA2},
\ee
\bea
\Delta\phi_z&=&\Delta\phi(\OM,\OL;z)\nonumber\\
&=& 2\ \tan^{-1}\left[ {
\sqrt{y_1-y_2}\ \ \left[\sqrt{y_3-\OM/(1-\OO)}-\sqrt{y_3-(1+z)\OM/(1-\OO)}\right]
\over
\sqrt{[y_1-\OM/(1-\OO)][y_2-(1+z)\OM/(1-\OO)]}+\sqrt{[y_1\longleftrightarrow y_2]}
}
\right],
\label{Del_phiA2}
\eea
where $y_1\longleftrightarrow y_2$ means repeat the previous term with $y_1$ 
and $y_2$ exchanged.

{\bf B.  Boundaries} 

1. $\OO\equiv\OM+\OL=1$

For the spatially flat model ($\ b\rightarrow \pm\infty)$
 a much simpler expression involving hypergeometric functions results:
\bea 
&&D_{\ell}(\OM,\OL=1-\OM,\nu=0;z)= {c\over H_0}(1+z)\int_0^z{dz\over \sqrt{1+\OM z(3+3z+z^2)}}
\nonumber\\
&& \hskip .5in
= {c\over H_0}{2(1+z)\over \OM^{1/3}}
\Biggr[
{}_2F_1\left( \frac16,\frac23;\frac76;1-\OM\right) 
\nonumber\\
&& \hskip .5in
- 
\left({1\over [1+\OM z(3+3z+z^2)]^{1/6} }\right)\,{}_2F_1\left( \frac16,\frac23;\frac76;{1-\OM\over 1+\OM z(3+3z+z^2)}\right)
\Biggl].
\label{2F1ansnu=0B1}
\eea
When $\OM\ne 1$ (\ref{2F1ansnu=0B1}) can be expressed as associated Legendre functions, 
\bea
&&D_{\ell}(\OM,\OL=1-\OM,\nu=0;z)= {c\over H_0}{
2^{1/6}\Gamma\left(1/6\right)(1+z)
\over 
3 [\OM^5(1-\OM)]^{1/12}
}\nonumber\\
&& \hskip 1in
\times\left[
{\rm P}^{-1/6}_{-1/6}\left({1\over \sqrt{\OM}}\right)-
{1\over (1+z)^{(1/4)}}\,{\rm P}^{-1/6}_{-1/6}\left(\sqrt{{1+\OM z(3+3z+z^2)}\over \OM(1+z)^3}\right)
\right].
\label{Pansnu=0B1}
\eea
If (\ref{Pansnu=0B1}) is given in terms of Legendre elliptic integrals the
result is more complicated:
\bea
D_{\ell}(\OM,\OL=1-\OM,\nu=0;z)
&&={c\over H_0}{1+z \over (3)^{1/4}\sqrt{\OM}(\OM^{-1}-1)^{1/6}}
\biggl[ 
-\{F(\phi_z,{\rm k})-F(\phi_0,{\rm k})\}
\biggr],
\nonumber\\
&&={c\over H_0}{1+z \over (3)^{1/4}\sqrt{\OM}(\OM^{-1}-1)^{1/6}}
\biggl[ 
-F(\Delta\phi_z,{\rm k})
\biggr],
\label{ansnu=0B1}
\eea
where
\be
{\rm k}^2=\left[{1\over 2}+{\sqrt{3}\over 4}\right],
\label{kOO=1}
\ee
\be
\phi_z=\phi(\OM;z)=\cos^{-1}
\left[
 1+z + (1-\sqrt{3})(\OM^{-1}-1)^{1/3}
\over
1+z + (1+\sqrt{3})(\OM^{-1}-1)^{1/3}
\right],
\label{phiOO=1}
\ee
and
\be
\Delta\phi_z=\Delta\phi(\OM,\OL;z)
\equiv2\ \tan^{-1}\left[{-z\ \sqrt{\sqrt{3}\OM(1/\OM-1)^{1/3}}
\sqrt{1+z[1+(1/\OM-1)^{1/3}]^{-1}}
\over
1+z [1+(1/\OM-1)^{1/3}]^{-1}+\sqrt{1+\OM z(3+3z+z^2)}}
\right].
\label{Del_phiOO=1}
\ee

2. $b=2$

This value of $b$ can be identified with ``critical" values of the cosmic 
parameters, \cite{FI}. We give a result good only for the lower $b=2$ curve, 
see (\ref{lower_b=con}). These models start with a big bang and 
expand to the the finite Einstein radius 
at $t=\infty$, see A3(vii-b) in the appendix of \cite{MG}:
\bea
&&D_{\ell}(\OM,\OL(\OM),\nu=0;z)= {c\over H_0} {1+z\over\sqrt{|1-\OO|}}
\nonumber\\
&& 
\times \sin\left\{ \ln\left(
{
\left[\sqrt{1/3-\Omega_m/(1-\Omega_0)} +1\right]
\left[\sqrt{1/3-(1+z)\Omega_m/(1-\Omega_0)} -1\right]
\over
\left[\sqrt{1/3-\Omega_m/(1-\Omega_0)} -1\right]
\left[\sqrt{1/3-(1+z)\Omega_m/(1-\Omega_0)} +1\right]
}
\right)
\right\}.
\eea

3. $\OL=0$

This result is due to \cite{MW}, we include it for completeness:
\be
D_{\ell}(\OM,\OL=0,\nu=0;z) ={2c\over H_0\OM^2}\left\{\OM
z+(\OM-2)\left(\sqrt{1+\OM z}-1\right)\right\}.  
\label{Mattig} 
\ee

4. $\OM=0$

These are massless big bang models, $\OL<1$, discussed by \cite{RH}:

\be
D_{\ell}(\OM=0,\OL,\nu=0;z)={c (1+z)\over H_0 \OL}
\left\{ 1+z-\sqrt{\OL+(1+z)^2(1-\OL)}\right\}.
\label{OM=0}
\ee

\vskip .25 in
\centerline{\bf II. $\nu=1$,\  66\% Filled-Beam Observations}
{\bf A. Four Open Big Bang Domains}
\bea 
&&D_{\ell}(\OM,\OL,\nu=1;z)=
\nonumber\\
&&{c\over H_0}2\,(1+z){\rm Sign}\left[3-\OM/(1-\OO)\right]
\sqrt{\left|
{
\left[3-\OM(1+z)/(1-\OO)\right]\left[3-\OM/(1-\OO)\right]
\over 
(1-\OO)[36+\OM^2\OL/(1-\OO)^3]
}\right|
}
\times \nonumber\\
&&
S_{(\OM,\OL,z)}
\Biggl[\ \sqrt{|(1-\OO)[36+\OM^2\OL/(1-\OO)^3]|}
\times \nonumber\\
&&
P\,\int_0^z{dz\over2\,\left[ 3-\OM(1+z)/(1-\OO) \right]\sqrt{(1+z)^2(1+\OM z)-z(z+2)\OL}}
\Biggr],
\label{ansnu=1A}
\eea
where 
\[
S_{(\OM,\OL,z)}[\ ]= \left\{
\begin{array}{l c  l}
{\rm cosh}[\ ] &:& b<0\,\&\,\left[3-\OM(1+z)/(1-\OO)\right]\left[3-\OM/(1-\OO)\right]<0, \\
{\rm sinh}[\ ] &:& b<0\,\&\,\left[3-\OM(1+z)/(1-\OO)\right]\left[3-\OM/(1-\OO)\right]>0, \\
{\rm sin}[\ ] &:& 0<b<486, \\
{\rm sinh}[\ ] &:& 486<b.
\end{array}
\right.
\]
Only the principal value of the integral (P) is needed and 
unlike the $\nu=0$ case, this integral takes on different forms when evaluated using 
Legendre elliptic integrals, depending on the 
value of the parameter $b$. Parts of the analytic result (\ref{ansnu=1A1})
sometimes diverge even though the 
total expression remains finite.
For example when $b=486$,  \ie when 
$\sqrt{36+\OM^2\OL/(1-\OO)^3}=0$ or equivalently $y_1=3$, a limit must be taken. 
The resulting $D_{\ell}$ on this new 
boundary can be found in 
II.B.5 below.  This new boundary splits the one open domain $2<b<\infty$ into two 
parts, see Fig.\,2. Consequently, the $\OM$--\ $\OL$ plane is more complicated 
for $\nu=1$ than for either $\nu=0$ or $\nu=2$. 
See A1 below for additional  trouble points that occur. 

{\bf 1.} For the three open domains defined by $b <  0, 2 < b<486$, and $486<b$ the luminosity distance $D_{\ell}$ takes the form:
\bea
&&D_{\ell}(\OM,\OL,\nu=1;z)=
\nonumber\\
&&{c\over H_0}2\,(1+z){\rm Sign}\left[3-\OM/(1-\OO)\right]
\sqrt{\left|
{
\left[3-\OM(1+z)/(1-\OO)\right]\left[3-\OM/(1-\OO)\right]
\over 
(1-\OO)[36+\OM^2\OL/(1-\OO)^3]
}\right|
}
\times \nonumber\\
&&
S_{(\OM,\OL,z)}
\Biggl[
{\kappa\ \sqrt{|36+\OM^2\OL/(1-\OO)^3|}
\over 2\,\sqrt{A}\ [A+\kappa(y_1-3)]}
\Biggl\{
\left[F(\phi_z,{\rm k})-F(\phi_0,{\rm k})\right]
\nonumber\\
&&
 +{A-\kappa(y_1-3)\over 2\kappa(y_1-3)}
\left[
{\rm P}\,\Pi(\phi_z,\hat{\alpha}^2,{\rm k})-{\rm P}\,\Pi(\phi_0,\hat{\alpha}^2,{\rm k})
\right]
\Biggr\}
+f_b
\Biggr],
\label{ansnu=1A1}
\eea

where $y_1,A,{\rm k},$ and $\phi_z$ are defined in (\ref{y1kappa})-(\ref{phiA1}) and the 
additional constant $\hat{\alpha}^2$ is:
\be
\hat{\alpha}^2\equiv {(A+\kappa(y_1-3))^2\over 4A\kappa(y_1-3)}.
\ee
$\Pi(\phi,\alpha^2,{\rm k})$ is the incomplete Legendre elliptic integral of the third 
kind\footnote{
$\Pi(\phi,\alpha^2,{\rm k})\equiv \int_0^{\phi} 1/\left[(1-\alpha^2\sin^2\phi)
\sqrt{1-{\rm k}^2\sin^2\phi}\ \right]\ d\phi$. In arriving at the results for the two-thirds filled 
beam model we discovered that equation 361.54 of \cite{BF} has the two square-root terms 
interchanged for the case $\alpha^2/(\alpha^2-1)> {\rm k}^2$.}  and P\,$\Pi(\phi,\alpha^2,{\rm k})$
is the principal part of that integral.
The function $f_b$ is one of,
\[ 
f_b = \left\{
\begin{array}{l c  l}
{1\over 4}\ln\left|\{[1+h(z)][1-h(0)]\}/\{[1+h(0)][1-h(z)]\}\right| &:& b<0\ {\rm or}\ 486<b, \\
{1\over 2}\left[\tan^{-1}h(z)-\tan^{-1}h(0)\right]&:& 2<b<486, 
\end{array}
\right. 
\]
where $h(z)$ is defined by:
\be
h(z)\equiv 
{
\sqrt{|36+\OM^2\OL/(1-\OO)^3|}
\sqrt{(1+z)\OM/|1-\OO|+\kappa y_1}
\over
(3-y_1)
\sqrt{
\left[(1+z)\OM/|1-\OO|-\kappa(1+y_1)/2\right]^2
-(1+y_1)(1-3y_1)/4
}
}.
\ee
Some care has to be taken when using these expressions. Divergences in the function $f_b$ 
necessarily occur and cancel divergences in  $\Pi(\phi,\alpha^2,{\rm k})$.
Divergences in  $f_b$ 
also occur which add to divergences in  $\Pi(\phi,\alpha^2,{\rm k})$ and cancel 
zeros in the multiplicative factor $\sqrt{\left|
\left[3-\OM(1+z)/(1-\OO)\right]\left[3-\OM/(1-\OO)\right]\right|}$ of (\ref{ansnu=1A1}).
Redshift independent divergences occur when $\OM/(1-\OO)=3$ and
when  $\OM(3-y_1)/(1-\OO)=y_1(2y_1+5)$. These points are plotted in Figure 2. 
Redshift dependent divergences occur at $(1+z)=3(1-\OO)/\OM$ and at 
$(1+z)\OM(3-y_1)/(1-\OO)=y_1(5+2y_1)$. These points appear in the $\OM$--\ $\OL$
plane respectively to the left of the $\OM/(1-\OO)=3$ line and between the 
$\OM(3-y_1)/(1-\OO)=y_1(5+2y_1)$ and $b=486$ curves.

Computer evaluation of (\ref{ansnu=1A1}) can be speeded up by reducing the number of
Legendre elliptic integrals that must be evaluated. As in (\ref{Del_ansnu=0}) we can
use the addition formula for $F(\phi,{\rm k})$, \ie
$F(\phi_z,{\rm k})-F(\phi_0,{\rm k})=F(\Delta\phi_z,{\rm k})$ and 
an addition formula for 
$\Pi(\phi,\alpha^2,{\rm k})$,\footnote{This equation is 116.03 of \cite{BF}, corrected
for two sign errors.}
\be
\Pi(\phi_z,\alpha^2,{\rm k})-\Pi(\phi_0,\alpha^2,{\rm k})=\Pi(\Delta\phi_z,\alpha^2,{\rm k})
+\frac12\sqrt{{\alpha^2\over (\alpha^2-1)(\alpha^2-{\rm k}^2)}}
\log
\left(
{1+
\xi
\over
1-
\xi
}
\right),
\label{AddFormulaPiA1}
\ee

where
\be
\xi\equiv {
\sin\phi_z\sin\phi_0\sin\Delta\phi_z\sqrt{\alpha^2(\alpha^2-1)(\alpha^2-{\rm k}^2)}
\over
1-\alpha^2\sin^2\Delta\phi_z-\alpha^2\sin\phi_z\sin\phi_0\cos\Delta\phi_z\sqrt{1-{\rm k}^2\sin^2\Delta\phi_z}
},
\ee
to cut the number of elliptic functions from four to two.
We were not able to simplify this expression enough to justify inclusion of a 
rewritten version of (\ref{ansnu=1A1}). However, it was used in our Fortran 
implementation (see Appendix).

{\bf 2.} For the open domain defined by $0 < b  < 2$ the luminosity distance $D_{\ell}$ has a somewhat simpler form:
\bea
&&D_{\ell}(\OM,\OL,\nu=1;z)=
{c\over H_0}{2\,(1+z)\over \sqrt{|1-\OO|}}
\sqrt{
{
\left[3-\OM(1+z)/(1-\OO)\right]\left[3-\OM/(1-\OO)\right]
\over 
\left[36+\OM^2\OL/(1-\OO)^3\right]
}
}
\times \nonumber\\
&&
{\rm sin}
\Biggl[
{
\sqrt{36+\OM^2\OL/(1-\OO)^3}
\over 
(3-y_1)\sqrt{y_1-y_2}
}
\Biggl\{
-\Biggl[F(\phi_z,{\rm k})-F(\phi_0,{\rm k})\Biggr]
\nonumber\\
&&\hskip .5in
+\left[
\Pi\left(\phi_z,{y_1-3\over y_1-y_2},{\rm k}\right)-\Pi\left(\phi_0,{y_1-3\over y_1-y_2},{\rm k}\right)\right]
\Biggr\}
\Biggr].
\label{ansnu=1A2}
\eea
The constants $y_1,\ y_2$ and ${\rm k}$, and the function $\phi_z$ are as defined in I.A.2 above
[see (\ref{y123})-(\ref{phiA2})].
Just as in the previous case, the number of Legendre elliptic functions in (\ref{ansnu=1A2})
can be reduced from four to two
by using the appropriate addition formulas. For $F(\phi,{\rm k})$ the formula is always 
the same, see (\ref{ansnu=0}) and (\ref{Del_ansnu=0}),
but because $\alpha^2$ is negative (\ref{AddFormulaPiA1}) changes to: 
\be
\Pi(\phi_z,\alpha^2,{\rm k})-\Pi(\phi_0,\alpha^2,{\rm k})=\Pi(\Delta\phi_z,\alpha^2,{\rm k})
-\frac12\sqrt{{\alpha^2\over (1-\alpha^2)(\alpha^2-{\rm k}^2)}}
\tan^{-1}
\left(\xi\right),
\label{AddFormulaPiA2}
\ee
where
\be
\xi\equiv {
\sin\phi_z\sin\phi_0\sin\Delta\phi_z\sqrt{\alpha^2(1-\alpha^2)(\alpha^2-{\rm k}^2)}
\over
1-\alpha^2\sin^2\Delta\phi_z-\alpha^2\sin\phi_z\sin\phi_0\cos\Delta\phi_z\sqrt{1-{\rm k}^2\sin^2\Delta\phi_z}
}.
\ee
This is 116.02 of \cite{BF} with one sign error corrected.

{\bf B.  Boundaries}

1. $\OO\equiv\OM+\OL=1$

For these models $\ b\rightarrow \pm\infty$ and 
 a much simpler expression results:
\bea 
&&D_{\ell}(\OM,\OL=1-\OM,\nu=1;z)
\nonumber\\
&& \hskip .5in 
= {c\over H_0}{2(1+z)^{3/2}\over \sqrt{1-\OM}}
{\rm sinh}\left[
{\sqrt{1-\OM}\over 2}\int_0^z{ dz\over (1+z) \sqrt{1+\OM z(3+3z+z^2)} }\right],
\nonumber\\
&& \hskip .5in
={c\over H_0\sqrt{1-\OM}}(1+z)^2
\Biggl[
\left(
{
1+\sqrt{1-\OM}
\over
\sqrt{1+\OM z(3+3z+z^2)} +\sqrt{1-\OM}
}
\right)^{ 1/3 }
\nonumber\\
&& \hskip 2in
-\left(
{
1-\sqrt{1-\OM}
\over
\sqrt{1+\OM z(3+3z+z^2)} -\sqrt{1-\OM}
}
\right)^{ 1/3}
\Biggr].
\label{ansnu=1FB}
\eea
This result can be given in terms of Legendre elliptic integrals $F(\phi,{\rm k})$
and $\Pi(\phi,\alpha^2,{\rm k})$; 
however, the authors can think of no useful purpose in doing so.

2. $b=2$

See the description for the $\nu=0$ case in section I.B.2 including (\ref{lower_b=con}) for this
``critical" value of $b$: 
\bea 
&&D_{\ell}(\OM,\OL(\OM),\nu=1;z)={c\over H_0}{(1+z)\sqrt{3/2}\over \sqrt{|1-\OO|}\,(11)}
\Biggl\{
\nonumber\\
&&
\sqrt{8}
\left[
\sqrt{1-3\,{\OM\over 1-\OO}}-
\sqrt{1-3\,{\OM(1+z)\over 1-\OO}}\
\right]
\cos\left({4\over \sqrt{6}}\log\,(h_z)\right)
\nonumber\\
&&
+\left[
8+\sqrt{\left(1-3\,{\OM\over 1-\OO}\right)\left(1-3\,{(1+z)\OM\over 1-\OO}\right)}\
\right]
\sin\left(
{4\over \sqrt{6}}\log\,(h_z)
\right)
\Biggr\},
\label{ansnu=1B2}
\eea
where $h_z$ is defined by:
\be
h_z=\left(
{
1+\sqrt{1/3-\OM/(1-\OO)}
\over
1+\sqrt{1/3-(1+z)\OM/(1-\OO)}
}
\right)
\sqrt{
{
2/3+(1+z)\OM/(1-\OO)
\over
2/3+\OM/(1-\OO)
}
}.
\ee
3. $\OL=0$

This result was first given by \cite{DC2},
\be
D_{\ell}(\OM,\OL=0,\nu=1;z) = {c\over H_0}{4\over
3\OM^2}\left[\left({3\over 2}\OM-1+{1\over 2}\OM z \right)\sqrt{1+\OM z} - \left({3\over 2}\OM-1\right)\right].
\label{DC2} 
\ee

4. $\OM=0$

This result is exactly the same as the $\nu=0$ result (\ref{OM=0}). If there is no mass
in the universe then removing 33\% of no mass from the beam changes nothing.

5. $b=486$

This result is equivalent to the $b\rightarrow 486$ limit of (\ref{ansnu=1A1})
but is simpler to use.
Because $\OL(\OM)$ is double valued for $b=$ constant $\ge 2$ , two expressions 
must be given to draw the $b=486$
curve, see Fig. 2.
For the upper part of the curve:
\be
\OL(\OM)=1-\OM+3\sqrt{2/b}\ \OM\ \cosh\left[{\cosh^{-1}
\left[\sqrt{b/2}\ (\OM^{-1}-1)\right]\over 3}\right],
\ee
where 
$0\le \OM\le 1/(1-\sqrt{2/b})$.
In this expression hyperbolic cosine analytically 
becomes cosine for $\OM\ge 1/(1+\sqrt{2/b})$. 
For the lower part of the curve:
\be
\OL(\OM)=1-\OM+3\sqrt{2/b}\ \OM\ \cos\left[{\cos^{-1}
\left[\sqrt{b/2}\ (1-\OM^{-1})\right]+\pi\over 3}\right],
\label{lower_b=con}
\ee
where 
$1\le \OM\le 1/(1-\sqrt{2/b})$.
The simplified result is:
\bea
&&D_{\ell}(\OM,\OL(\OM),\nu=1;z)
={c\over H_0}{(1+z)\over \sqrt{|1-\OO|}(33)^{(3/4)}}
\sqrt{
\left[3-{\OM(1+z)\over (1-\OO)}\right]\left[3-{\OM\over (1-\OO)}\right] 
}
\times 
\nonumber\\
&&
\Biggl\{
F\left(\phi_0,{\rm {\sqrt{33}+5\over 2\sqrt{33}}}\right)-F\left(\phi_z,{\rm {\sqrt{33}+5\over 2\sqrt{33}}}\right)
-2\left[E\left(\phi_0,{\rm {\sqrt{33}+5\over 2\sqrt{33}}}\right)-E\left(\phi_z,{\rm {\sqrt{33}+5\over 2\sqrt{33}}}\right)\right]
\nonumber\\
&&
+2\,(33)^{(1/4)}
\Biggl[
{\sqrt{8+\left[2+\OM/(1-\OO)\right]^2}
\over
\sqrt{3-\OM/(1-\OO)}\left[3+\sqrt{33}-\OM/(1-\OO)\right]
}
\nonumber\\
&&
-\,{\sqrt{8+\left[2+(1+z)\OM/(1-\OO)\right]^2}
\over
\sqrt{3-(1+z)\OM/(1-\OO)}\left[3+\sqrt{33}-(1+z)\OM/(1-\OO)\right]
}
\Biggr]
\Biggr\}.
\label{ansnu=1B486}
\eea
The arguments of the elliptic functions, $\phi_z$ and $\phi_0$, can be calculated
from (\ref{phiA1}) using $y_1=3$ and $A=\sqrt{33}$. To reduce the number of 
elliptic functions needed to evaluate (\ref{ansnu=1B486}), addition formulas for 
$F(\phi,{\rm k})$ and $E(\phi,{\rm k})$ can be used 
[see (\ref{Del_ansnu=0}), (\ref{AddFormulaE}), and (\ref{AddFormulaE_A1})]. 
The value of $\Delta\phi_z$ is given by (\ref{Del_phiA1}).

\vskip .25 in
\centerline{\bf III. $\nu=2$, Empty-Beam Observations}
{\bf A. Three Open Big Bang Domains}
\bea 
D_{\ell}(\OM,\OL,\nu=2;z)&=&{c\over H_0}(1+z)^2
\int_0^z{dz\over(1+z)^2\sqrt{(1+z)^2(1+\OM z)-z(z+2)\OL}}.
\label{nu=2}
\eea
Like the $\nu=1$ case this integral takes on different forms when evaluated in terms of 
Legendre elliptic integrals, depending on the 
value of the parameter $b$. 
  
{\bf 1.} For the two open domains defined by $b < 0$ and $2 < b$ the luminosity distance $D_{\ell}$ takes the form:
\bea 
D_{\ell}(\OM,\OL,\nu=2;z)&=&{c\over H_0}{(1+z)^2\over \OL}
\Biggl\{
\nonumber\\
&&
\hskip -1.75in
-(A+\kappa y_1)
\Biggl[
{
\sqrt{(1+z)^2(1+\OM z)-z(z+2)\OL}
\over 
(1+z)[(1+z)\OM/|1-\OO|+A+\kappa y_1]
}
-{1\over \OM/|1-\OO|+A+\kappa y_1}
\Biggr]
\nonumber\\
&&
\hskip -1.75in -
{
(A-\kappa y_1)\sqrt{|1-\OO|}
\over
2\sqrt{A}
}
\Biggl[
F(\phi_z,{\rm k})-F(\phi_0,{\rm k})
\Biggr]
+\sqrt{A}\sqrt{|1-\OO|}
\Biggl[
E(\phi_z,{\rm k})-E(\phi_0,{\rm k})
\Biggr]
\Biggr\}
\label{ansnu=2A1}
\eea
where $y_1,A,{\rm k},$ and $\phi_z$ are defined in (\ref{y1kappa})-(\ref{phiA1}).\footnote{
$E(\phi,{\rm k})\equiv \int_0^{\phi} \sqrt{1-{\rm k}^2\sin^2\phi}\ d\phi$}
Just as with the result for the $\nu=1$ case, \ie (\ref{ansnu=1A1}), the number of 
Legendre elliptic integrals required to evaluate (\ref{ansnu=2A1}) can be reduced from
four to two by using addition formulas 116.01 of \cite{BF}. The addition formula
for $E(\phi,{\rm k})$ is:
\be
E(\phi_z,{\rm k})-E(\phi_0,{\rm k})=E(\Delta\phi_z,{\rm k})
-{\rm k}^2\sin\phi_z\sin\phi_0\sin\Delta\phi_z.
\label{AddFormulaE}
\ee
For this case 
\bea
&&-{\rm k}^2\sin\phi_z\sin\phi_0\sin\Delta\phi_z=
-{2\left[2A+\kappa(1+3y_1)\right]\over\left[(1+z)\OM/(1-\OO)-y_1-\kappa A\right]}
\nonumber\\
&&\times
{ \sqrt{ \left[(1+z)\OM/(1-\OO)-y_1\right]\left[ \OM/(1-\OO)-y_1\right] }\over
\left[\OM/(1-\OO)-y_1-\kappa A\right] 
\left[\tan(\Delta\phi_z/2)+1/\tan(\Delta\phi_z/2)\right]
},
\label{AddFormulaE_A1}
\eea
where an expression for $\tan(\Delta\phi_z/2)$ is given by (\ref{Del_phiA1}). 
  
{\bf 2.} For the domain $0 < b < 2$ the luminosity distance $D_{\ell}$ takes the form:
\bea 
&&D_{\ell}(\OM,\OL,\nu=2;z)=
\nonumber\\
&&
\hskip .25in
{c\over H_0}{(1+z)^2\over \OL}
\Biggl\{
-y_3
\Biggl[
{
\sqrt{(1+z)^2(1+\OM z)-z(z+2)\OL}
\over 
(1+z)[(1+z)\OM/|1-\OO|+y_3]
}
-{1\over \OM/|1-\OO|+y_3}
\Biggr]
\nonumber\\
&&
\hskip .25in -
{y_2\sqrt{|1-\OO|}
\over
\sqrt{y_1-y_2}
}
\Biggl[
F(\phi_z,{\rm k})-F(\phi_0,{\rm k})
\Biggr]-
\sqrt{y_1-y_2}\sqrt{|1-\OO|}
\Biggl[
E(\phi_z,{\rm k})-E(\phi_0,{\rm k})
\Biggr]
\Biggr\},
\label{ansnu=2A2}
\eea
where the constants $y_1,\ y_2,\ y_3$  and ${\rm k}$  are defined in (\ref{y123})-(\ref{k+})
but the function $\phi_z$ is now defined as 
\bea
\phi_z=\phi(\OM,\OL;z)
&=&\sin^{-1}\sqrt{
{
(1+z)\OM/|1-\OO|+y_2
\over
(1+z)\OM/|1-\OO|+y_3
}
}.
\label{phi++}
\eea
For this case the value of $\Delta\phi_z$ needed to reduce the number of elliptic integrals is
the NEGATIVE of that given by (\ref{Del_phiA2}) for the $\nu=0$ case.
When the addition formula (\ref{AddFormulaE}) is used, an
additional term is contributed to (\ref{ansnu=2A2}) which can be evaluated using,
\bea
&&-{\rm k}^2\sin\phi_z\sin\phi_0\sin\Delta\phi_z=
{\OM(y_1-y_3)|1-\OO|^{(-3/2)}(y_1-y_2)^{(-1/2)}\over [(1+z)\OM^2/(1-\OO)^2-(2+z)y_1\OM/(1-\OO)-2y_1(1+y_1)]}
\nonumber\\
&&
\times
\left\{{[y_2-\OM/(1-\OO)]\sqrt{(1+z)^2(1+\OM z)-z(z+2)\OL}\over [y_3-(1+z)\OM/(1-\OO)]}-
{[y_2-(1+z)\OM/(1-\OO)]\over [y_3-\OM/(1-\OO)]}\right\}.
\eea
{\bf B.  Boundaries} 

1. $\OO\equiv\OM+\OL=1$

This case is the $\ b\rightarrow \pm\infty$ limit of (\ref{nu=2}) and 
 a simpler expression containing hypergeometric functions results:
\bea 
&&D_{\ell}(\OM,\OL=1-\OM,\nu=2;z)
\nonumber\\
&& 
\hskip .5in
={c\over H_0}(1+z)^2
\Biggl\{
1-{1\over (1+z)\sqrt{1+\OM z(3+3z+z^2)}}
\nonumber\\
&& 
\hskip 1in
+
\frac35\ \OM^{1/3}\Biggl[
\left( 
{1 \over[1+\OM z(3+3z+z^2)]^{5/6} }\right)\,
{}_2F_1\left( \frac56,\frac13;\frac{11}6;{1-\OM\over 1+\OM z(3+3z+z^2)}\right)
\nonumber\\
&& 
\hskip 2in
-{}_2F_1\left( \frac56,\frac13;\frac{11}6;1-\OM\right) 
\Biggr]
\Biggr\}.
\label{2F1ansnu=2B1}
\eea
When $\OM\ne 1$, (\ref{2F1ansnu=2B1}) can be expressed in terms of 
associated Legendre functions as,
\bea 
&&D_{\ell}(\OM,\OL=1-\OM,\nu=2;z)=
\nonumber\\
&& 
\hskip .5in{c\over H_0}(1+z)^2
\Biggl\{
1-{1\over (1+z)\sqrt{1+\OM z(3+3z+z^2)}}
+
{\Gamma\left(5/6\right)\over 2^{1/6}}
\left[{\OM\over 
1-\OM}
\right]^{5/12}
\nonumber\\
&& 
\times
\left[
{(1+z)^{1/4}\over 
\sqrt{1+\OM z(3+3z+z^2)}}
{\rm P}^{-5/6}_{1/6}\left(\sqrt{{1+\OM z(3+3z+z^2)}\over \OM(1+z)^3}\right)
-{\rm P}^{-5/6}_{1/6}\left({1\over \sqrt{\OM}}\right)
\right]
\Biggr\}
.
\label{Pansnu=2B1}
\eea
When $\OM\ne 1$, (\ref{2F1ansnu=2B1}) can also be expressed in terms of  
Legendre elliptic integrals as,
\bea 
D_{\ell}(\OM,\OL=1-\OM,\nu=2;z)&=&{c\over H_0}{(1+z)^2\over 1- \OM}
\Biggl\{
\nonumber\\
&&
\hskip -2.5in
-(\sqrt{3}+1)\left(\OM^{-1}-1\right)^{1/3}
\Biggl[
{
\sqrt{1+\OM z(3+3z+z^2)}
\over 
(1+z)[1+z+(\sqrt{3}+1)\left(\OM^{-1}-1\right)^{1/3}]
}
-{1\over 1+(\sqrt{3}+1)\left(\OM^{-1}-1\right)^{1/3}}
\Biggr]
\nonumber\\
&&
\hskip -1.5in -
{1\over (\sqrt{3}+1)(3)^{1/4}}
\sqrt{\OM}\left(\OM^{-1}-1\right)^{1/6}
\Biggl[
F(\phi_z,{\rm k})-F(\phi_0,{\rm k})
\Biggr]
\nonumber\\
&&
\hskip -1.0in +
(3)^{1/4}\sqrt{\OM}\left(\OM^{-1}-1\right)^{1/6}
\Biggl[
E(\phi_z,{\rm k})-E(\phi_0,{\rm k})
\Biggr]
\Biggr\},
\label{ansnu=2B1}
\eea
where the constant ${\rm k}$ is given by (\ref{kOO=1}) and the functions $\phi_z$ 
and $\Delta\phi_z$ are given respectively by (\ref{phiOO=1}) and (\ref{Del_phiOO=1}). 
For this case the additional term needed to use the addition formula (\ref{AddFormulaE}) 
in (\ref{ansnu=2B1}) is:
\bea
&&-{\rm k}^2\sin\phi_z\sin\phi_0\sin\Delta\phi_z
\nonumber\\
&&
={
z\ 2(3)^{3/4}\left(2+\sqrt{3}\,\right)\sqrt{1-\OM}
\over 
\left[1+z+\left(1+\sqrt{3}\,\right)\left(\OM^{-1}-1\right)^{1/3}\right]
\left[1+\left(1+\sqrt{3}\,\right)\left(\OM^{-1}-1\right)^{1/3}\right]
}
\nonumber\\
&&
\times
{
\left\{z+ \left[1+\left(\OM^{-1}-1\right)^{1/3}\right]\left[1+\sqrt{1+\OM z(3+3z+z^2)}\right]\right\}
\over
\left\{2+3\,z\,\OM+z^2\,\OM\left[1+\left(1+\sqrt{3}\,\right)\left(\OM^{-1}-1\right)^{1/3}\right]
+2\sqrt{1+\OM z(3+3z+z^2)}\right\}
}.
\label{AddFormulaE_B1}
\eea
2. $b=2$

See the description for the $\nu=0$ case in section I.B.2 including (\ref{lower_b=con}) for this
``critical" value of $b$: 
\bea 
&&D_{\ell}(\OM,\OL(\OM),\nu=2;z)
\nonumber\\
&&={c\over H_0}{9\,\OM\,(1+z)^2 \over 2|1-\OO|^{3/2}}
\Biggl\{
{1\over (1+z)}\sqrt{\frac13-{(1+z)\OM\over 1-\OO}} -\sqrt{\frac13-{\OM\over 1-\OO}} 
\nonumber\\
&&
+{\OM\over 1-\OO}
\log\left[
{
1+\sqrt{1/3-(1+z)\OM/(1-\OO)}
\over
1+\sqrt{1/3-\OM/(1-\OO)}
}
\sqrt
{
{
2/3+\OM/(1-\OO)
\over
2/3+(1+z)\OM/(1-\OO)
}
}
\right]
\Biggr\}.
\eea

3. $\OL=0$

This result was first given by \cite{DC1},
\bea
&&D_{\ell}(\OM,\OL=0,\nu=2;z) \nonumber\\ &&\hskip .5 in ={c\over H_0}{\OM(1+z)^2\over
4(1-\OM)^{3/2}}\Biggl[{3\OM\over 2(1-\OM)} \ln\left\{ \left({1+\sqrt{1-\OM}\over1-\sqrt{1-\OM}
}\right) \left({\sqrt{1+\OM z}-\sqrt{1-\OM} \over \sqrt{1+\OM z}+\sqrt{1-\OM} }\right)
\right\}\nonumber\\ &&\hskip .5 in +{3\over\sqrt{1-\OM}}\left({\sqrt{1+\OM z}\over 1+z} -1\right)
+{2\sqrt{1-\OM}\over \OM}\left( 1 - {\sqrt{1+\OM z}\over (1+z)^2}\right)\Biggr]\,, 
 \label{DC1}
\eea 
and can be rewritten using the identity 
\be
\sinh^{-1}\sqrt{{1-\OM\over \OM(1+z)}} =
{1\over2} \ln\left({\sqrt{1+\OM z}+\sqrt{1-\OM} \over \sqrt{1+\OM z}-\sqrt{1-\OM} }\right)\,. 
\ee
 When $\OM > 1$ equation (\ref{DC1})
is analytically continued using $\sqrt{1-\OM} \longrightarrow \pm i \sqrt{\OM-1}$, which
simplifies by using, $\sinh^{-1}(i x) = i \sin^{-1}(x)$ to give a form containing only real
variables. The $\OM = 1$ result for all $\nu$ was given by \cite{DV}: 
\be
D_{\ell}(\OM=1,\OL=0,\nu;z) = {c\over H_0}{1\over
(\nu+{1\over2})}\left[(1+z)^{({\nu\over2}+1)}- (1+z)^{(-{\nu\over2}+{1\over 2})}\right].
\label{Dash} 
\ee
4. $\OM=0$

This result is exactly the same as the $\nu=0$ and $\nu=1$ result (\ref{OM=0}). If there is no mass
in the universe then removing 100\% of no mass from the beam removes nothing.

\section{Conclusions} \label{sec-conclusions}

We have given useful forms for the luminosity distance in three currently relevant 
cosmologies. They are all dynamically FLRW cosmologies in the large but 
differ in how gravitating
matter effects optical observations. The models are labeled by an additional 
parameter $\nu$ ($\nu$ = 0, 1, and 2) beyond the familiar $H_0,\Omega_m,$ and $\Lambda$. The 
$\nu=0$ model is standard FLRW where all matter is homogeneous and transparent 
on the scale 
of the observing beam widths. This model is called the `filled-beam' model.
The 
$\nu=2$ model  assumes the opposite; all matter is inhomogeneous and excluded 
from the observing beams. 
This extreme case is called the `empty-beam' model. 
The 
$\nu=1$ model assumes that 1/3 of the mass density of the universe is excluded 
from observing beams and hence it is the `two-thirds filled-beam' model.
These three cases were singled out because their distance-redshift relations
can be given in terms of incomplete elliptic integrals; functions which are universally 
available in computer libraries and very efficiently evaluated.\footnote{
The results appearing in Section \ref{sec-results} have been coded 
and are posted at http://www.nhn.ou.edu$\sim$thomas/z2dl.html.
This code is discussed in the Appendix and compared to the numerical 
integration times of \cite{KHS}.} 
For the $\nu=1$ and 2 cases, somewhat simpler expressions than what we have given exist, but 
only for complex arguments of the elliptic integrals. We chose to give 
expressions whose arguments are real and which 
can be rapidly evaluated.
Results are available 
for all $0\le\nu\le 2$ but only in terms of the less familiar and unavailable 
Heun functions, \cite{KRa}. We have extended the flat space, $\OO=1$, results given 
here  to arbitrary filling parameter $\nu$. These new results will be available shortly. 
Related results have been independantly found by \cite{DM}.
A calculation similar to the $\nu=1$ case given here is that 
of the age of the Universe as a 
function of redshift and can be found in \cite{TK}.

\acknowledgements
R. Kantowski wishes to thank VP for Research, E. Smith, for funds
to support J.K. Kao's visit to OU during the summer of 1998 when the first elliptic 
integral results were obtained. R. C. Thomas thanks P. Helbig for 
discussions of his code, see
\cite{KHS}, and E. Baron for benchmarking discussions. 

\appendix
\section{Appendix} 
\label{sec-appendix}

One expected practical use of the results given in this paper is to
speedup distance evaluations for the $\nu=$ 0,1,2  partially filled beam FLRW models.
 We have implemented and made publicly available a Fortran 90 version 
of this work called Z2DL (see http://www.nhn.ou.edu$\sim$thomas/z2dl.html 
for Z2DL with documentation and extensive
CPU-time benchmark results).
Z2DL uses Carlson elliptic integrals (see \cite{PTVF}
and references therein)
and results in a fast distance calculator. 
We have benchmarked Z2DL by comparing it with the commonly used and 
fast numerical integration 
routine ANGSIZ (see \cite{KHS}).  For a given
($\OM,\OL$), the total CPU-time required to convert $5\times 10^5$ redshifts (equally
spaced between z=0 and z=5) to luminosity distance using Z2DL and ANGSIZ 
separately were
recorded.   By calculating the ratio of ANGSIZ CPU-time to Z2DL CPU-time 
on a grid of points in ($\OM,\OL$) we have generated three speedup surfaces,
one for each value of $\nu=$ 0,1,2 (see Fig. 3 for the $\nu=0$ surface). The results 
for all three comparisons are given as contour plots at the web site.
Using an IBM AIX 375 MHz Power III
approximately 7 hours was required to generate each
($\OM,\OL$) grid of 30 x 30 points (minus models without a big bang).

For the purpose of a clearer presentation, we omitted speedup points along
the $\OM=0$ and $\OL=0$ lines.  Along these boundaries
speedup factors are greater than 100. The large open domains of the 
 $\Omega_m$-$\OL$ plane, \ie subsection `A' cases, 
constitute the majority of
models in the grid and also those with the least impressive speedup.
However, even for these cases, the improvement is substantial: typically  
17-20 for $\nu=0$ (standard filled beam FLRW), 6-8 for $\nu=1$ (66\%
filled beam FLRW), and  11-13 for $\nu=2$ (empty beam FLRW).

To gauge the level of agreement between distances computed by
ANGSIZ and Z2DL, a finer grid of ($\OM,\OL$) with 3000 x 3000 points
(between 0 and 3 in both directions, also excluding models without a
big bang) was used.  For each ($\OM,\OL$), both routines were used to
compute luminosity distance for z=1.  Most often the results agree to within
one part in $10^6$.  Cases where disagreements greater than one part in
$10^3$ occur are near the upper b=2 line (see Fig. 1). We found that 
ANGSIZ was giving less accurate distances near this boundary of non-big bang models
as ANGSIZ documentation explains.

\clearpage

\figcaption[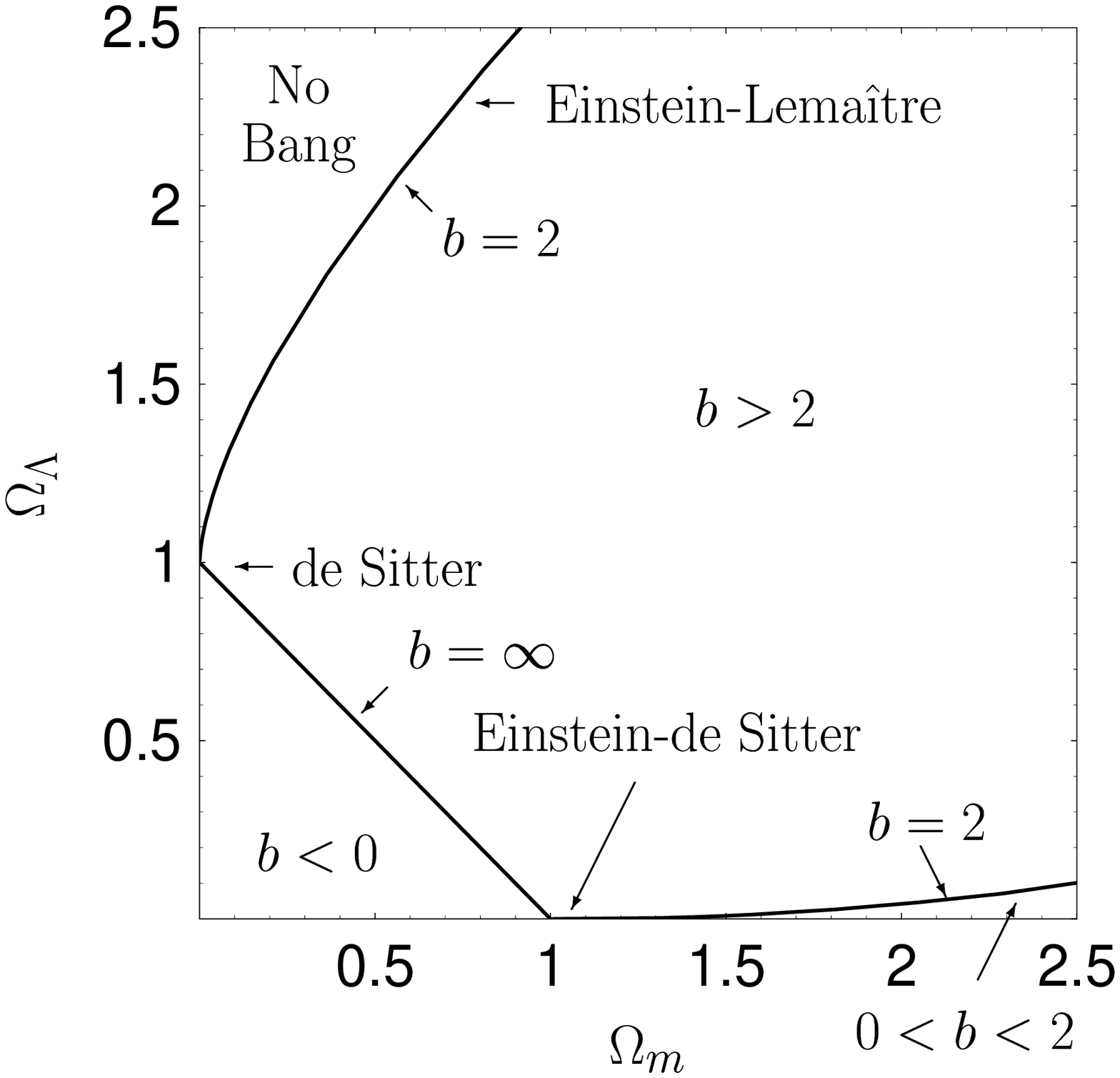]{The $\Omega_m$-$\OL$ plane showing various $b$ domains 
that require different expressions for distance-redshift $D_{\ell}$ for all
three cases: $\nu=0,1,2$ \ie filled-beam, 66\% filled-beam, and empty-beam. 
\label{fig1_3}}

\figcaption[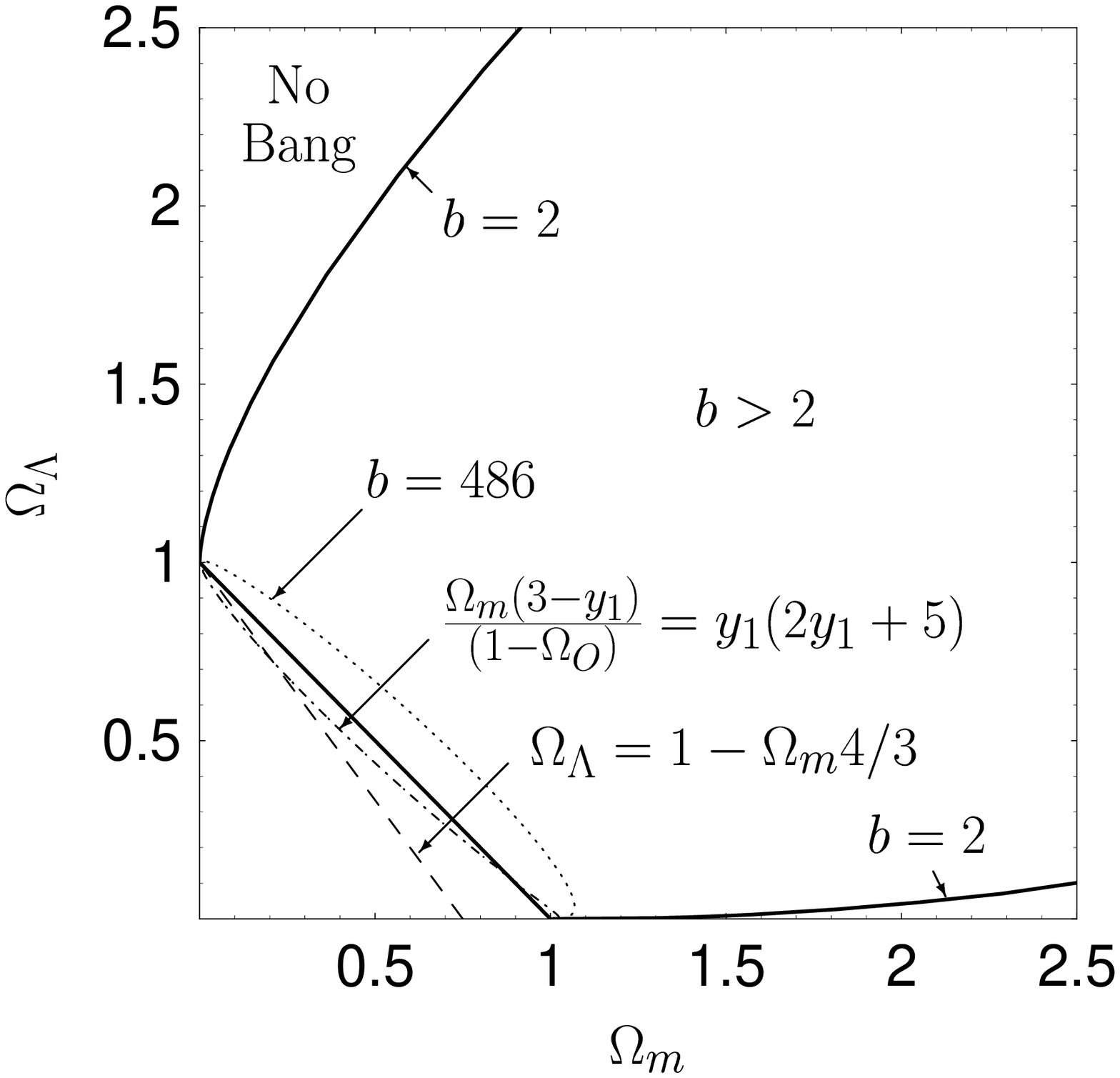]{Additional domains in the $\Omega_m$-$\OL$ plane 
for $\nu=1$, \ie for 66\% filled-beam observations, where complications due 
to divergent terms occur in the analytic results.  
For $\OM$--\ $\OL$ values on the dashed and
dot-dashed lines, define respectively by $\OL=1-\OM4/3$ 
and  $\OM(3-y_1)/(1-\OO)=y_1(2y_1+5)$, expression (\ref{ansnu=1A1}) must be evaluated 
by taking  a numerical limit. For points to the left of the straight dashed line and points 
between the dot-dashed and $b=486$ curves, a single value of  $z$ exits for which 
(\ref{ansnu=1A1}) also diverges. These $z$ values are defined respectively
by  $(z+1)=3(1-\OO)/\OM$ and 
$(1+z)\OM(3-y_1)/(1-\OO)=y_1(5+2y_1)$.  For $\OM,\OL$, and $z$ satisfying either
equation  a limiting process must be used to evaluate 
$D_{\ell}$ via (\ref{ansnu=1A1}), see the Appendix. 
For points on the divergent $b=486$ curve an analytic limit 
was obtained in (\ref{ansnu=1B486}).
\label{fig2_3}}

\figcaption[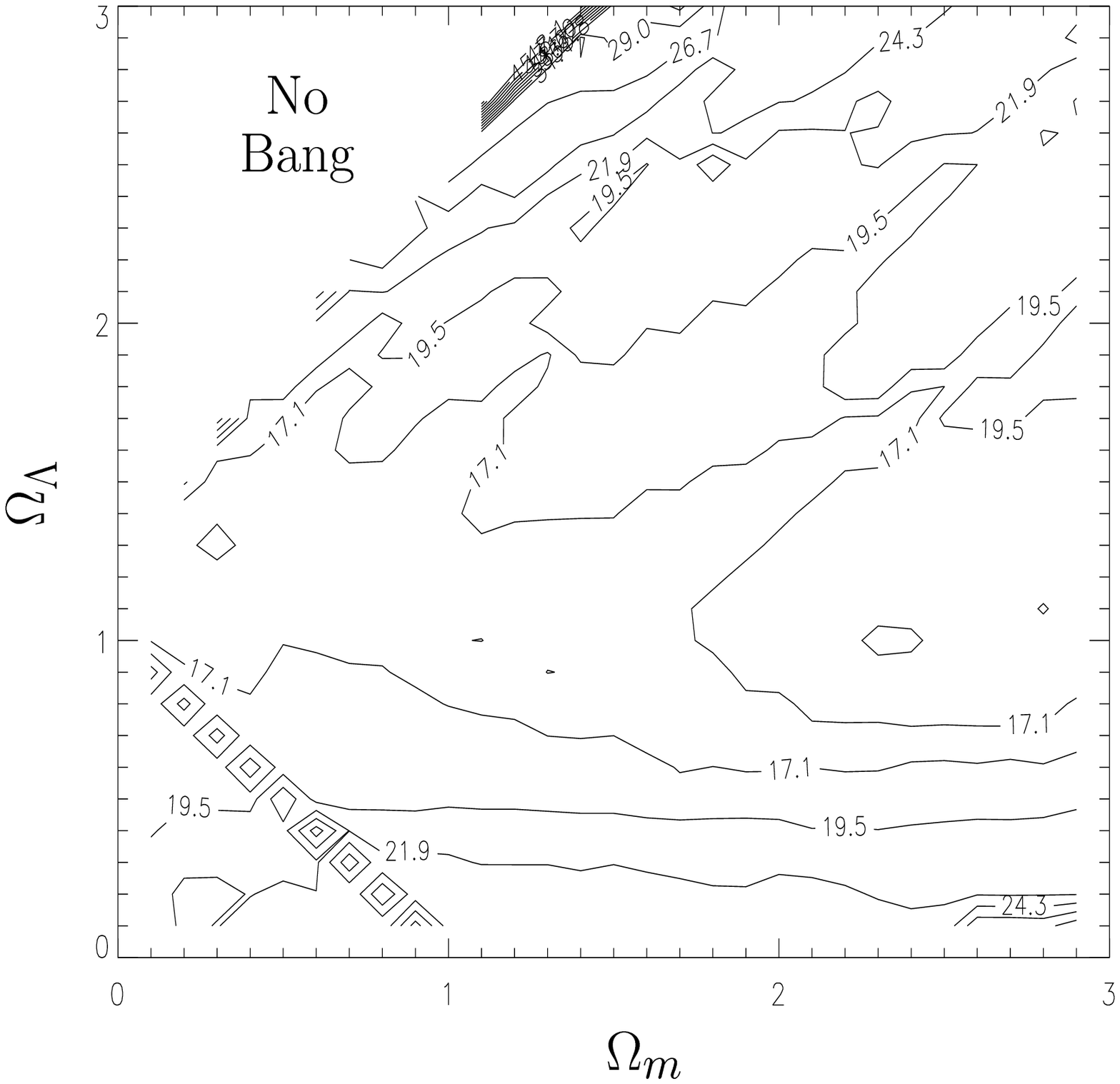]{Contour plot of the $\Omega_m$-$\OL$ plane showing speedup 
factors for  Z2DL over ANGSIZ when $\nu=0$ (standard filled beam FLRW cosmology). 
Speedup factors for the other
two cases considered in this paper, $\nu=1,2$ \ie the 66\% filled-beam and empty-beam 
can be found at the web site. 
\label{fig3_3}}

\plotone{fig1_3.eps}
\eject
\plotone{fig2_3.eps}
\eject
\plotone{fig3_3.eps}

\end{document}